\documentclass[reprint,superscriptaddress,showkeys]{revtex4-2}
\usepackage{verbatim}
\usepackage{color}
\usepackage{subfigure}
\usepackage[]{graphicx}
\usepackage{dcolumn}
\usepackage{bm}
\usepackage{blindtext}
\usepackage{floatrow}
\usepackage{makeidx}
\usepackage{epsfig}
\usepackage{hyperref}
\usepackage{array}
\usepackage{epstopdf}
\usepackage{braket}
\usepackage{multirow}
\usepackage{mathtools}
\usepackage{amsmath,amssymb,amsfonts}
\usepackage{fancyhdr}
\usepackage{multirow}
\newcommand{\RNum}[1]
{\uppercase\expandafter{\romannumeral #1\relax}}
\newcommand{\Rnum}[1]
{\lowercase\expandafter{\romannumeral #1\relax}}
\def\qe{\textsc{Quantum ESPRESSO}}

\def\bz{Brillouin zone}
\makeindex
\begin{document}
\preprint{APS/123-QED}
\title{Electronic structure of molybdenene from first principles}
\author{Ashkan Shekaari}
\email{shekaari@email.kntu.ac.ir}
\email{shekaari.theory@gmail.com}
\affiliation{Department of Physics, K. N. Toosi University of Technology, Tehran 15875-4416, Iran}
\author{Mahmoud Jafari}
\email{jafari@kntu.ac.ir}
\affiliation{Department of Physics, K. N. Toosi University of Technology, Tehran 15875-4416, Iran}
\date{\today}
\begin{abstract}
Density functional theory has been applied to investigate the electronic structure and lattice stability of molybdenene monolayer in both its hexagonal and triclinic phases, within ultrasoft pseudopotential approach. In agreement with experimental findings, it has been found that either phase is metallic. Analyzing partial density of states has revealed that the $d$ valence orbitals of molybdenum atoms have had the largest contribution to such a metallic property, due to being half-empty as well as being the outermost. Phonon-dispersion calculations also have led to negative frequencies for either phase, showing lattice instability, as reported in the experimental literature.\\
\end{abstract}
\maketitle
\section{\label{sec:1}INTRODUCTION}
Beginning with the discovery of graphene~\cite{0,gap} about two decades ago, many areas of science and technology, from materials science~\cite{mt} to electronics~\cite{el} to medicine~\cite{nm}, have been revolutionized since then, making researchers look for other possible low-dimensional materials with similar uniqueness. Some paramount of such materials investigated so far are two-dimensional (2D) MXenes~\cite{mzen}, MoS$_2$~\cite{ms1,ms2}, WS$_2$~\cite{ws2}, phagraphene~\cite{pg}, borophene~\cite{b36}, SLSiN~\cite{sl} (single-layer silicon nitride), ﬂuorographenes~\cite{2x,3x}, graphene oxides~\cite{4x}, B$_2$C graphene sheet~\cite{5x}, h-BN~\cite{6x}, Ti$_2$C and Ti$_3$C~\cite{11x}, and so on~\cite{son}.\\

Characterized by massless quasi-particles in their low-energy excitation spectrum, Dirac materials~\cite{1} also fall into the category of low-dimensional nanoscale materials, which include $d$-wave superconductors~\cite{2}, Weyl and Dirac semi-metals, graphene, topological insulators~\cite{3}, and superfluid phases of $^3$He~\cite{4}. They exhibit a range of unique and universal properties: the same power-law temperature dependence of fermionic specific heat~\cite{1}, response magnetic fields and impurities, transport properties, suppressed back-scattering, and optical conductivity~\cite{6}.\\

In contrast to other kinds of 2D materials, elemental Dirac or Dirac-like materials (such as phosphorene~\cite{ph}, graphene, silicene~\cite{si}, borophene, and 2D gold~\cite{2g}) are of especial interest, for they typically lack structural impurities, leading to enhanced electron mobility as a result. Advancements in discovery of low-dimensional and Dirac materials have further led to the inclusion of higher-atomic-number metallic atoms such as Mo, W, and Ti, the electronic densities of which have to be distributed only in two dimensions. They result in potential exotic electronic and excitonic properties, which, say, make them mechanically robust under high loads and temperatures~\cite{bk}.\\

As a common behavior, none of the Dirac materials ever identified has so far showed metallic properties. A recent investigation has experimentally confirmed the formation of free-standing molybdenene monolayer~\cite{7} as a 2D Dirac material with metallic properties nevertheless, which is purely composed of Mo atoms. According to the literature, during the production of molybdenene, weakly-bonded molybdenene sheets are first formed, which, later on, exhibit metallic properties upon exfoliation. This so-called molybdenene monolayer, with electrical conductivity of about 940 S.m$^{-1}$, has also been reported to acquire tunable electronic and optical properties upon hybridization with 2D h-BN and MoS$_2$. Functionalized molybdenene monolayers with different elements including H, Li, Be, B, C, N, O, F, Na, Mg, Al, Si, P, S, and Cl at both basal surfaces~\cite{baz}, as well as the electronic structure and anomalous superconducting properties of Li/F modified molybdenene have also recently been investigated theoretically~\cite{sup}.\\

Experimentally, the two phases (eigenstructures) of molybdenene have so far been identified at different areas of the monolayer, taking hexagonal and triclinic crystal systems (Fig.~\ref{fig:2}), abbreviated here as HM (hexagonal molybdenene) and TM (triclinic molybdenene). The electronic structure of molybdenene monolayer has formerly been reported computationally~\cite{7} within the framework of projector-augmented waves (PAW)~\cite{paw}. Nevertheless, we here have adopted a different approach based on ultrasoft pseudopotentials (USPP)~\cite{11} to investigate the metallic properties and lattice dynamics of either eigenstructure, and demonstrate that similar results will be found.\\

The present work has been organized as follows: Sec.~\ref{sec:2} describes computational details, Sec.~\ref{sec:3} discusses the results obtained; and Sec.~\ref{sec:4} concludes our discussion.
\section{\label{sec:2}Computational details}
\begin{sloppypar}
	We have adopted the self-consistent~\cite{10}, plane-wave, ultrasoft, pseudopotential approach of density functional theory (DFT)~\cite{8}, at the PBE-GGA~\cite{12} level of approximation, as implemented in \qe\ (v.7.0) integrated suite~\cite{13,131,132}. The running mode has been multi-core parallelized by Open MPI~\cite{mpi}, on Linux (Debian) operating system~\cite{deb,lin}. The scalar-relativistic, ultrasoft~\cite{14} pseudopotential file {{\texttt{Mo.pbe-spn-rrkjus\_psl.0.2.UPF}}}~\cite{15}, generated by Rappe-Rabe-Kaxiras-Joannopoulos (RRKJ)~\cite{16} pseudization recipe with non-linear core correction~\cite{17}, has been used to model the core electrons; the valence shell has also included $4s^24p^65s^14d^55p^0$ orbitals according to that file. The electronic density has been augmented by means of a Fourier interpolation scheme in real space~\cite{18}. The kinetic energy cut-off values of about 1224 and 4900 eV have been found to be optimized for wavefunctions, and for the charge density and the potential, respectively. Two primitive cells including one hexagonal (for HM) and one triclinic (for TM) have been used, each containing two Mo atoms (Fig.~\ref{fig:2}), under periodic boundary conditions, with optimized vacuum lengths of $>15$ \AA\ along the $z-$axis to eliminate periodic interactions. The initial lattice constant (before relaxation) for either phase has been set to 3.70 \AA\ (the experimental value reported in the literature~\cite{7}). The equilibrium lattice constant ($a_0$) of each phase has been calculated via applying a variable-cell relaxation based on Murnaghan isothermal equation of state~\cite{41,42} as implemented in thermo\_pw (1.7.0) driver~\cite{thermo}, which applies a uniform strain to the unit cell; a fixed-cell relaxation has also been applied using the obtained lattice constants to find the exact atomic positions. The self-consistency energy convergence threshold has been set to $10^{-12}$ eV. A force convergence threshold of about $2.57\times 10^{-5}$ eV.\AA$^{-1}$ for ionic relaxation based on quasi-Newtonian Broyden-Fletcher-Goldfarb-Shanno (BFGS) geometry optimization method~\cite{19} has been used. A threshold of about $10^{-12}$ has also been applied for phonon-dispersion self-consistency. The \bz\ samplings of HM and TM have been carried out along $\Gamma$--M--K--$\Gamma$, and $\Gamma$--Y--S--X--$\Gamma$, meshed by about 100 k-points, respectively; the smearing method has also been set to Methfessel-Paxton~\cite{20}.\\
\end{sloppypar}
All the values/schemes reported/applied here have been determined/adopted optimally based on their associated energy-convergence tests. All the software packages used in the present work are also either free under or compatible with the GNU general public license (GPL)~\cite{gpl}.\\
\section{\label{sec:3}Results and discussion}
\subsection{Crystal structure}
We have used the Murnaghan isothermal equation of state to find the equilibrium, zero-pressure lattice constants of the two monolayers. This approach assumes a linear response for the bulk modulus $K_0$ of a solid with respect to a change in pressure (i.e., a finite strain), leading to
\begin{equation}
	\label{eq:pv}
	P(V)=\frac{K_{0}}{K'_{0}}\left[\left(\frac{V_0}{V}\right)^{K'_{0}}-1\right],
\end{equation}
where $K'_{0}$ is the first derivative of the bulk modulus with respect to pressure $P$, $V$ is volume of the unit cell, and $V_0$ is the zero-pressure volume. The constancy of $K$ and $K'$ guarantees the linear dependence of the bulk modulus on pressure. Integrating Eq.~\eqref{eq:pv} using $P(V)=-dE(V)/dV$ (the first law of thermodynamics at absolute zero) results in
\begin{equation}
	\label{eq:uv}
	E(V)=E_{gs}+\frac{{K_{0}}V}{K'_{0}}\left[\frac{(V_{0}/V)^{K'_{0}}}{{K'_{0}}-1}+1\right]-\frac{{K_{0}}V_0}{{K'_{0}}-1},
\end{equation} 
where $E$ is the internal energy and $E_{gs}$ is the ground-state energy of the system. After relaxing the atomic positions, a uniform strain is applied to change the lattice constant with an increment of about $\pm 2.65 \times 10^{-2}$ {\AA}. The total energy of the system is then estimated for each strained structure using self-consistent-field calculations. Fitting Eq.~\eqref{eq:uv} to the calculated energy values results in $K_0$ and $K'_0$, as tabulated in Table~\ref{tab:0}.\\
\begin{widetext}
\begin{minipage}{\linewidth} 
\begin{table}[H]
	\caption{\label{tab:0}
		The values of bulk modulus $K_0$ and its first derivative $K'_0$, equilibrium volume $V_0$ and lattice constant $a_0$, ground-state energy $E_{gs}$, mass $M$ (in atomic mass unit a.m.u.), and density $\rho$, obtained using Murnaghan equation of state.}
	\begin{tabular}{c|c|c|c|c|c|c|c}
		\hline
		Phase&$K_0$ (GPa)&$K'_0$&$V_0$ ({\AA$^3$})&$a_0 $(\AA)&$E_{gs}$ (eV)& M (amu)&$\rho$ (gr/cm$^3$)\\
		\hline		
		HM&8.9&4.62&253.47&4.1820&-4012.840&191.920&1.2573\\
		\hline
		TM&119.3&15.0&243.15&3.8090&-4012.079&191.920&1.3107\\
		\hline\\
	\end{tabular}
\end{table}
\end{minipage}
\end{widetext}
The isothermal $P-V$ (Eq.~\ref{eq:pv}) and $E-V$ (Eq.~\ref{eq:uv}) diagrams could then be obtained for each phase, as illustrated in Fig.~\ref{fig:1}.
\begin{widetext}
\begin{minipage}{\linewidth} 
\begin{figure}[H]
	\centering
	\subfigure[HM]{\label{subfig:1(a)}
		\includegraphics[scale=0.29]{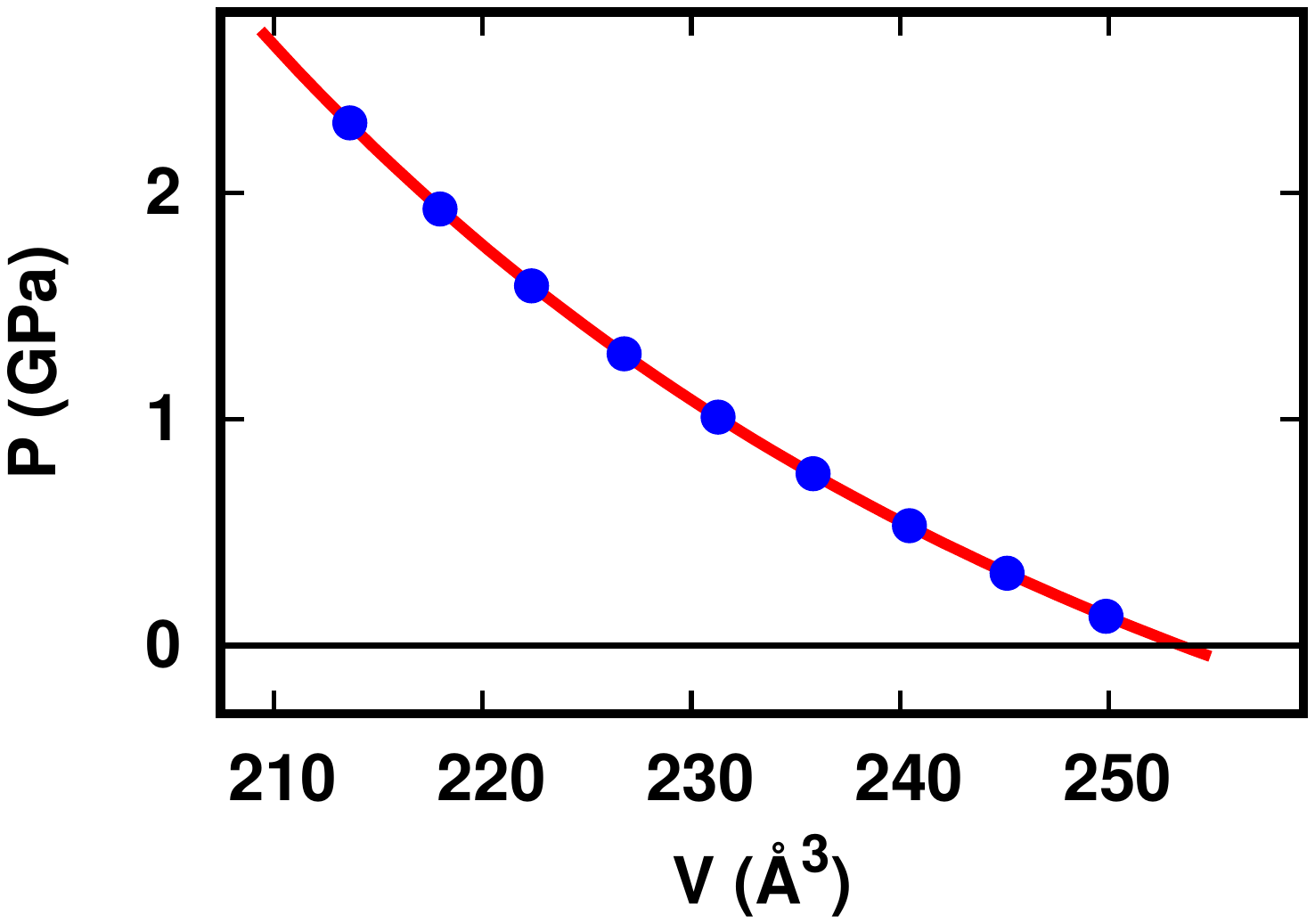}}
	\subfigure[HM]{\label{subfig:1(b)}
		\includegraphics[scale=0.29]{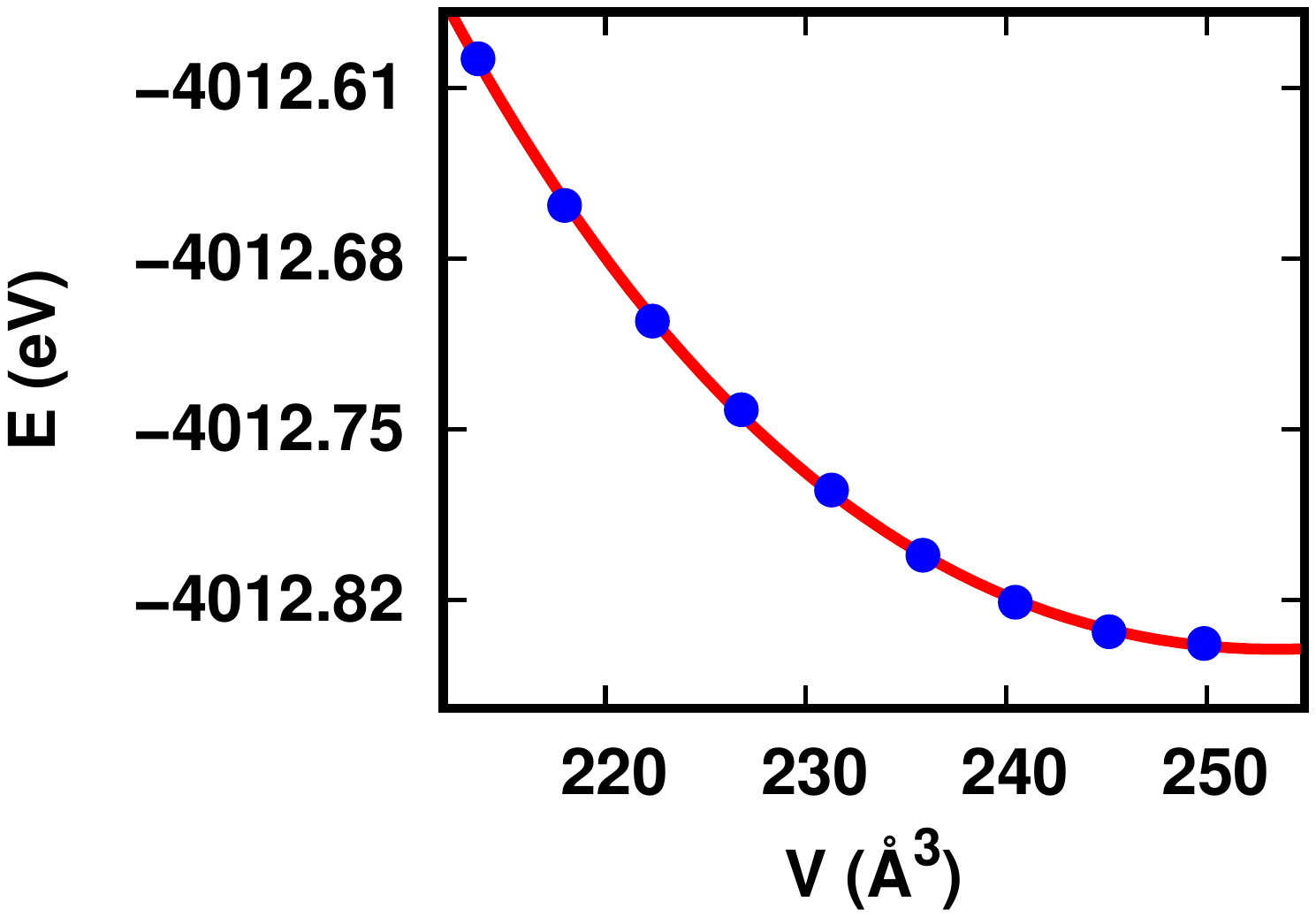}}
	\subfigure[TM]{\label{subfig:1(c)}
		\includegraphics[scale=0.29]{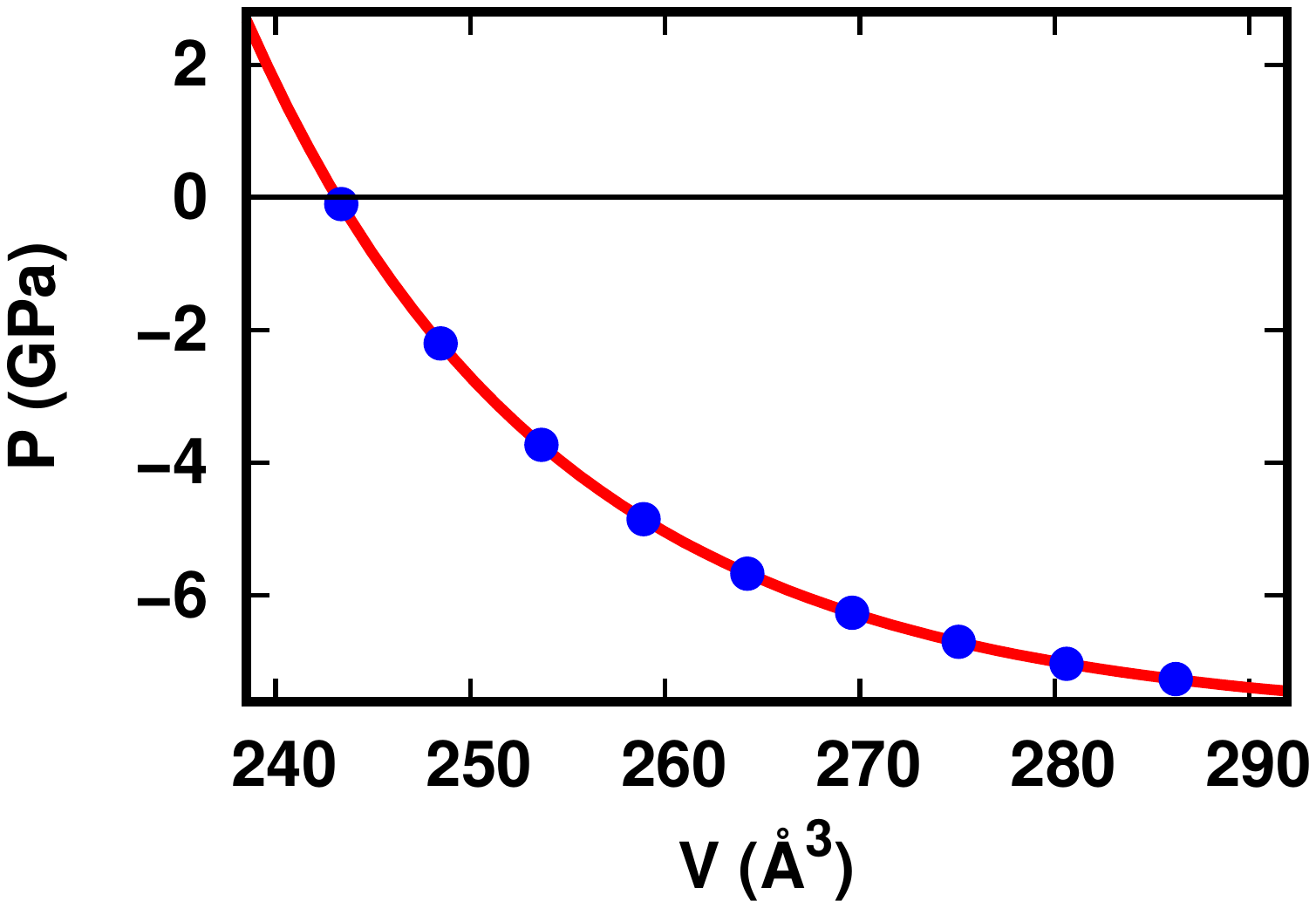}}
	\subfigure[TM]{\label{subfig:1(d)}
		\includegraphics[scale=0.29]{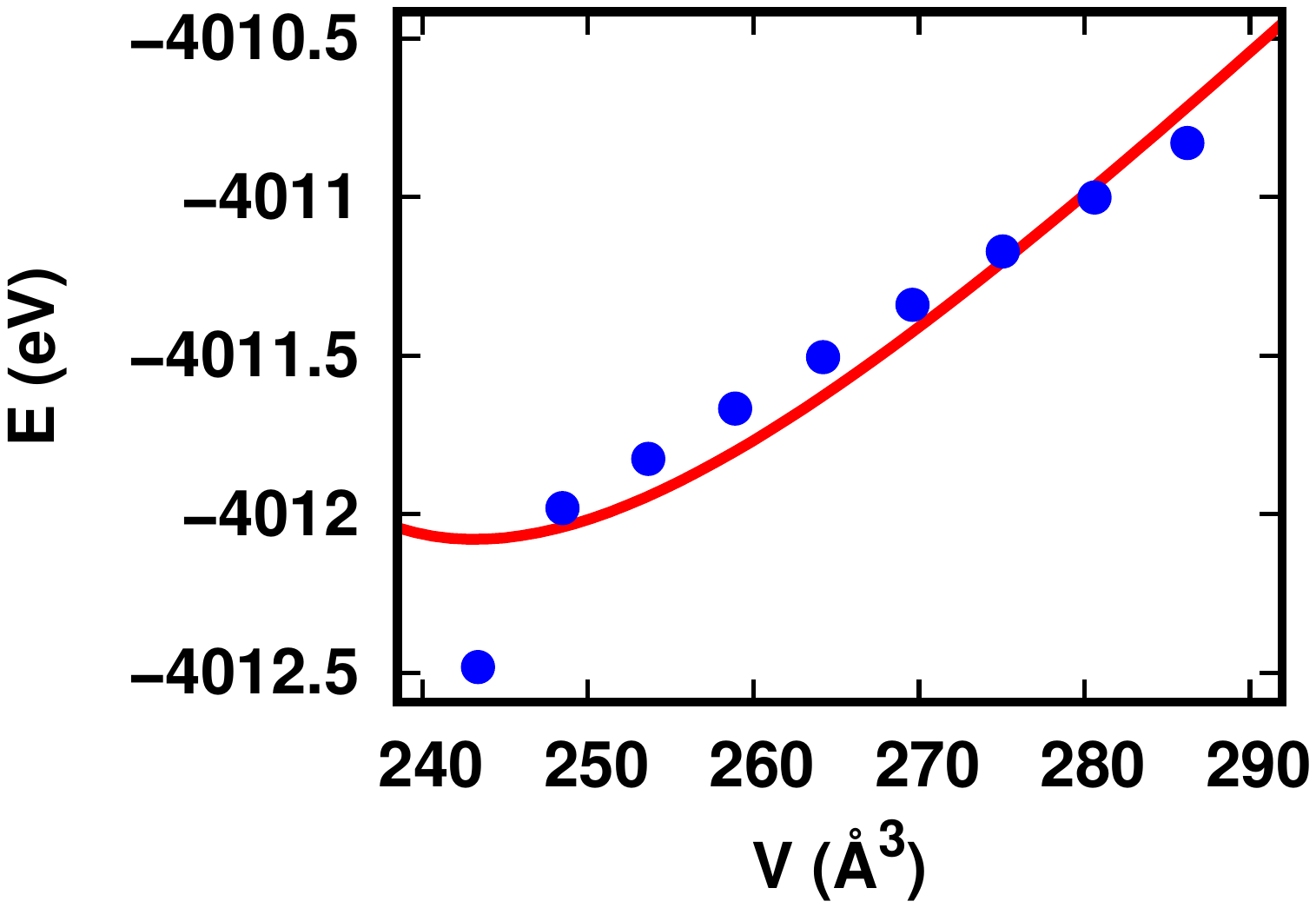}}	
	\caption{\label{fig:1}
		The $P-V$ [(a) and (c)] and $E-V$ [(b) and (d)] diagrams for HM [(a) and (b)] and TM [(c) and (d)] monolayers, obtained using Eqs.~\ref{eq:pv} and \ref{eq:uv}---rendered in Gnuplot~\cite{gnup}. The zeros of the $P-V$ diagrams correspond with the minimums of the $E-V$ diagrams (namely, $E_{gs}$).}
\end{figure}
\end{minipage}
\end{widetext}
As is seen, the minimums of the $E-V$ curves indicate the ground-state energies $E_{gs}$ of the monolayers. They also correspond with the points at which the pressure becomes zero. Fig.~\ref{fig:2} illustrates the relaxed atomic structures of the two phases of molybdenene.
\begin{widetext}
\begin{minipage}{\linewidth} 
\begin{figure}[H]
	\centering
	\subfigure[HM]{\label{subfig:2(a)}
		\includegraphics[scale=0.35]{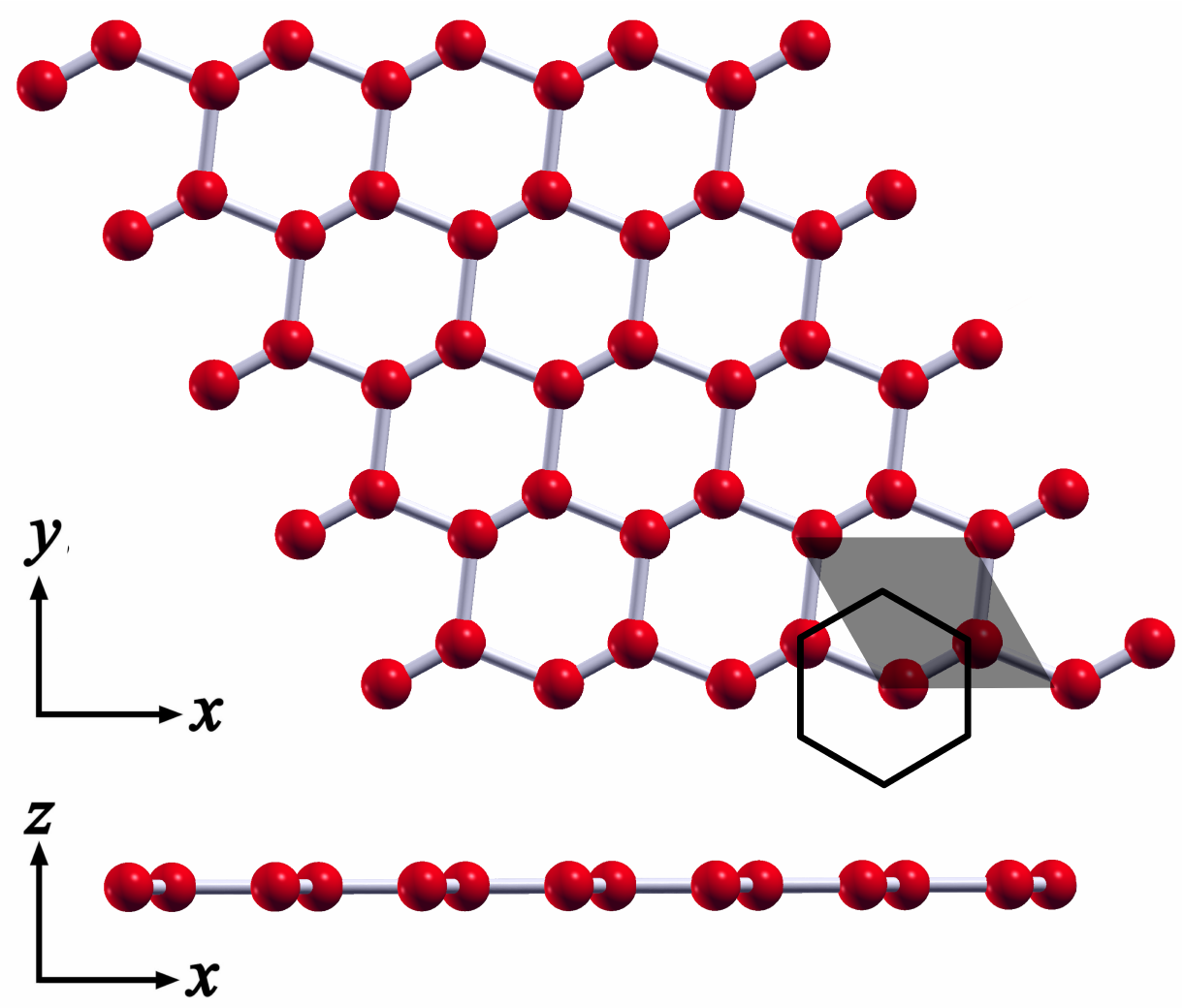}}
	\subfigure[TM]{\label{subfig:2(b)}
		\includegraphics[scale=0.47]{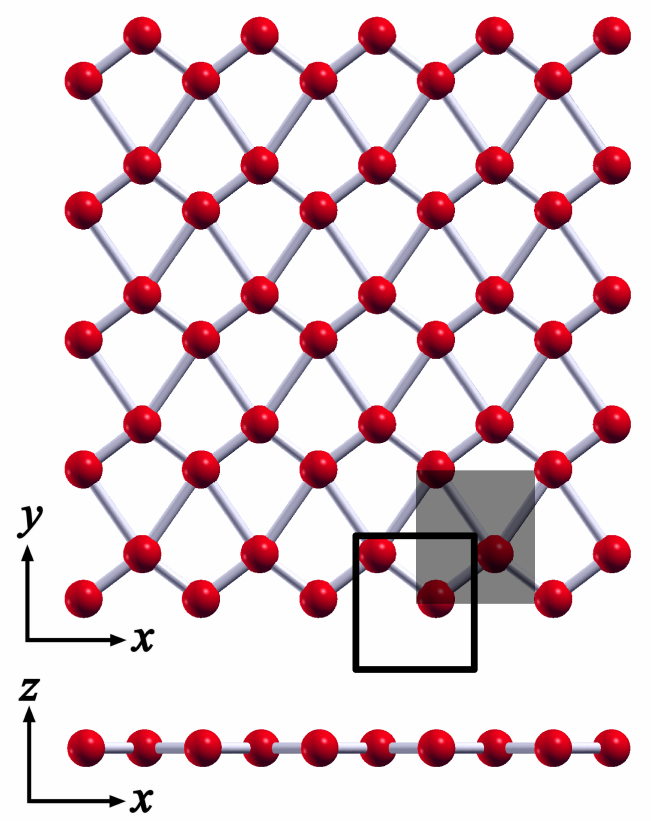}}	
	\caption{\label{fig:2}
		The $5\times 5$ supercells of (a) HM and (b) TM in $x-y$ and $x-z$ planes. The associated conventional unit cells (gray areas) and Wigner-Seitz unit cells (thin black-border areas) have also been illustrated--rendered in XCrySDen~\cite{xc}. Either unit cell contains two Mo atoms.}
\end{figure}
\end{minipage}
\end{widetext}
After relaxation, the Mo--Mo bond length between the two atoms of the unit cell has been found as 2.0517 \AA\ for HM, and 2.3938 \AA\ for TM. The relaxed atomic positions have also been tabulated in Table.~\ref{tab:1}.
\begin{table}[H]
	\caption{\label{tab:1}
		Cartesian components of the atomic positions of HM and TM after relaxation.}
	\begin{tabular}{c|cccc}
		\hline
		Phase & Atom&$x$ (\AA) & $y$ (\AA) & $z$ (\AA)\\
		\hline
		\multirow{2}{*}{HM}&1&0.4013 & 0.2599 & 0.0000\\
		&2&2.2021 & 1.2432 & 0.0000\\
		\hline
		\multirow{2}{*}{TM} &1&0.5487 &0.0236 & 0.0000\\
		&2&2.4513s & 1.4764 & 0.0000\\
		\hline
	\end{tabular}
\end{table}
\subsection{Electronic structure}
Fig.~\ref{fig:3} illustrates the electronic band structures of HM and TM. As is seen, the energy bands cross the Fermi level (zero of the diagrams), indicating the metallic property of either phase, in agreement with the literature.
\begin{widetext}
\begin{minipage}{\linewidth} 
\begin{figure}[H]
	\centering
	\subfigure[HM]{\label{subfig:3(a)}
		\includegraphics[scale=0.3]{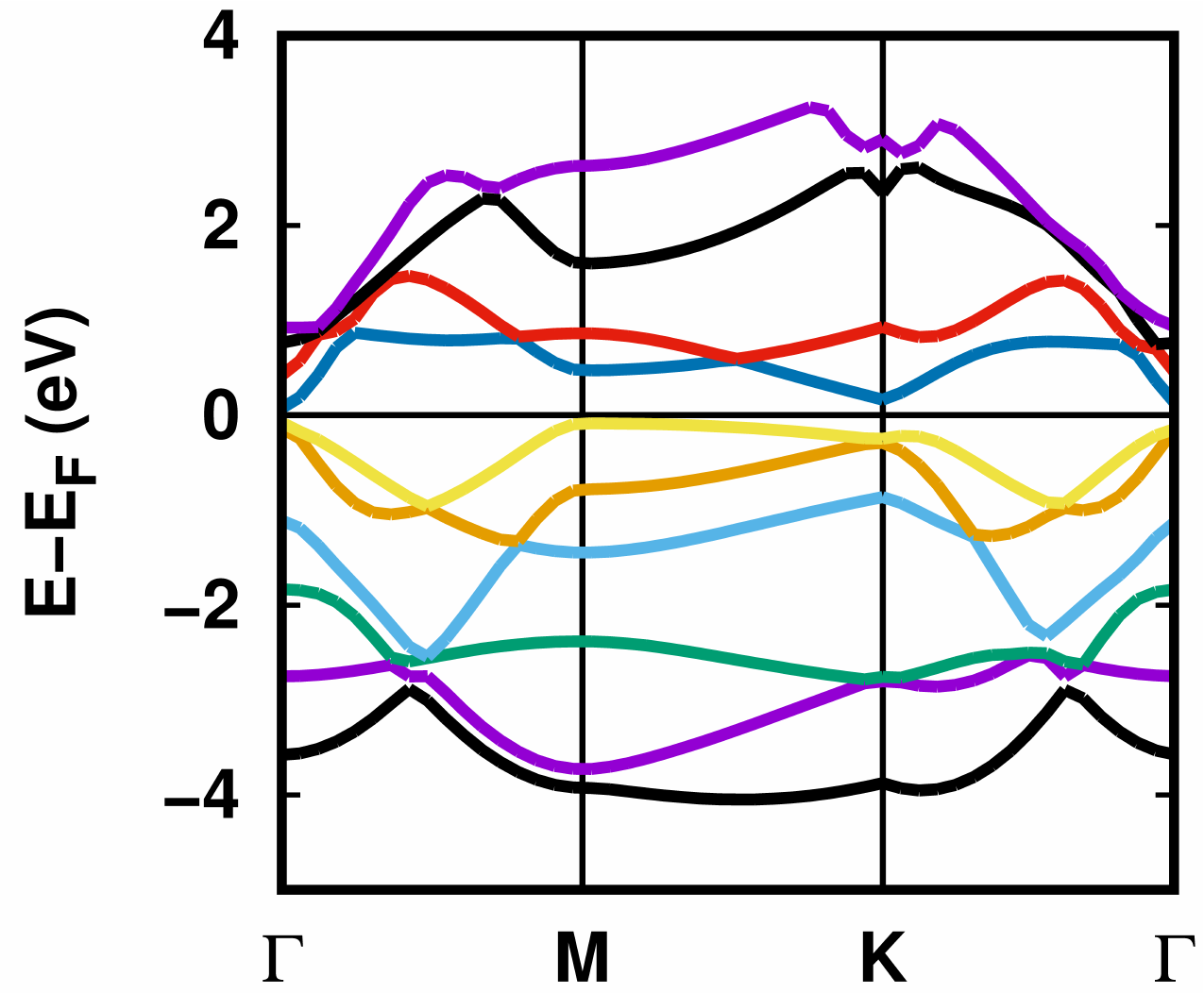}}
	\subfigure[TM]{\label{subfig:3(b)}
		\includegraphics[scale=0.3]{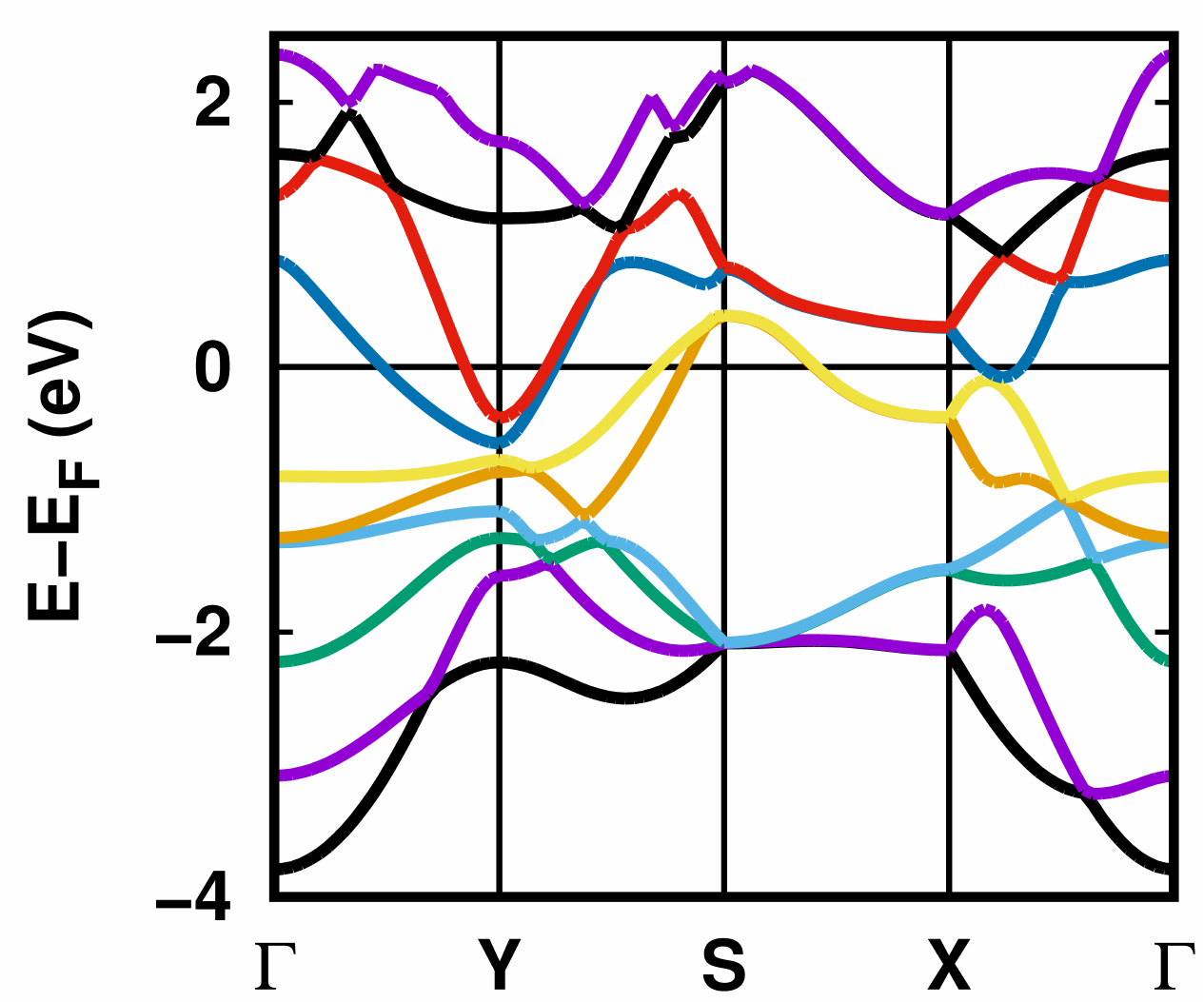}}	
	\caption{\label{fig:3}
		Electronic band structures of (a) HM and (b) TM. The Fermi energy is zero of the diagrams, being crossed by the energy bands as an indication of metallic feature.}
\end{figure}
\end{minipage}
\end{widetext}

\begin{widetext}
\begin{minipage}{\linewidth}
\begin{figure}[H]
	\centering
	\subfigure[HM]{\label{subfig:4(a)}
		\includegraphics[scale=0.29]{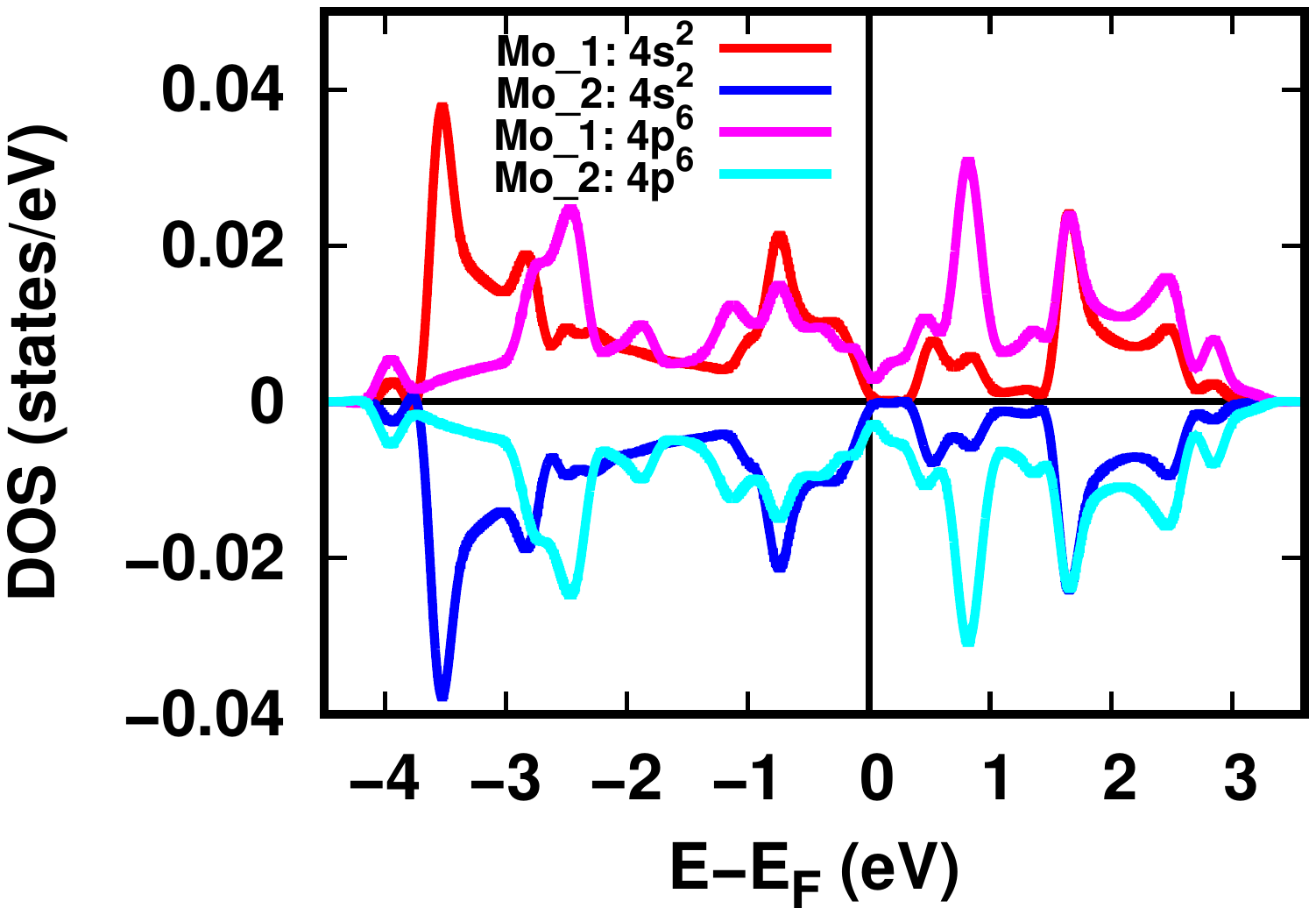}}
	\subfigure[HM]{\label{subfig:4(b)}
		\includegraphics[scale=0.29]{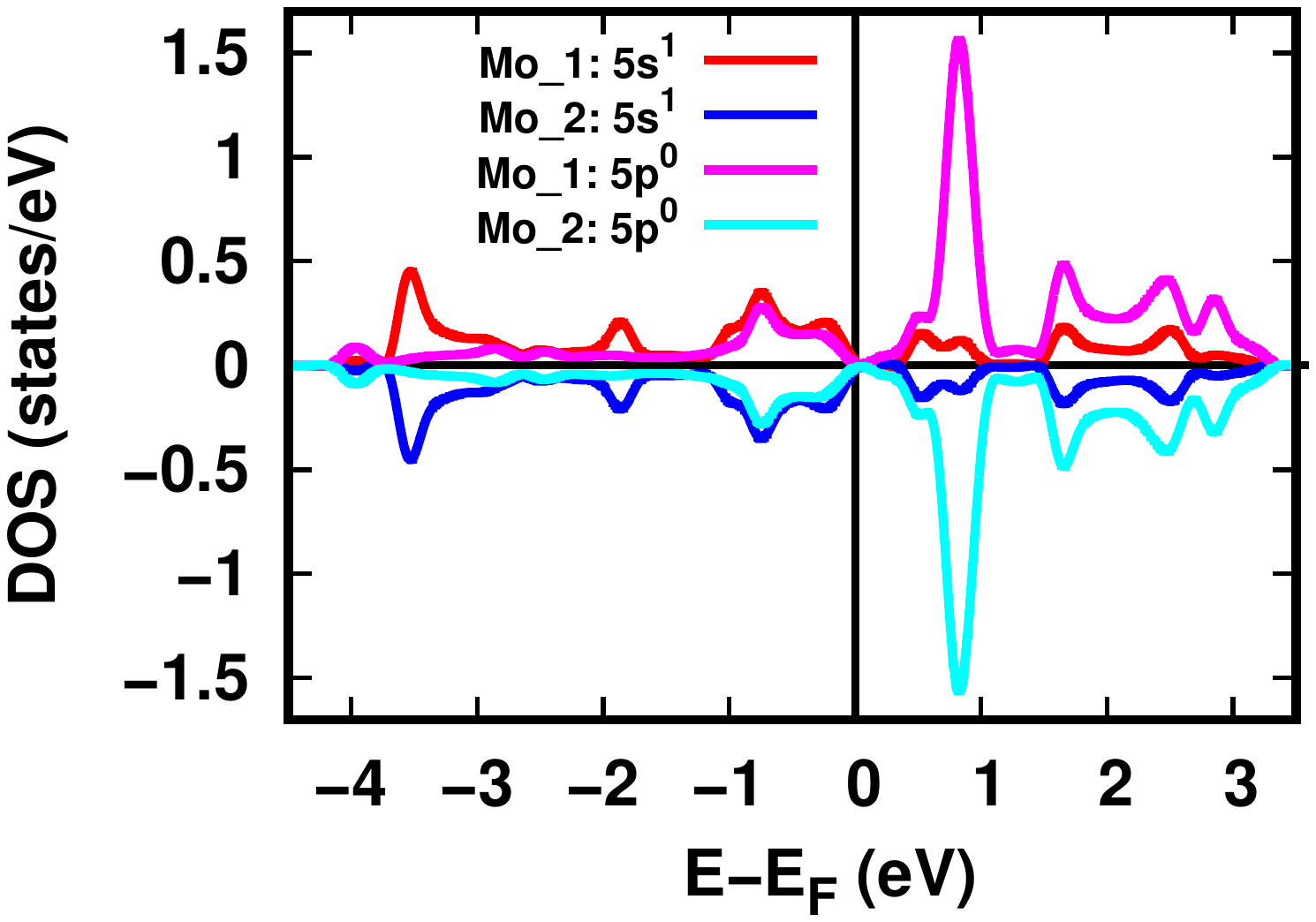}}
	\subfigure[HM]{\label{subfig:4(c)}
		\includegraphics[scale=0.29]{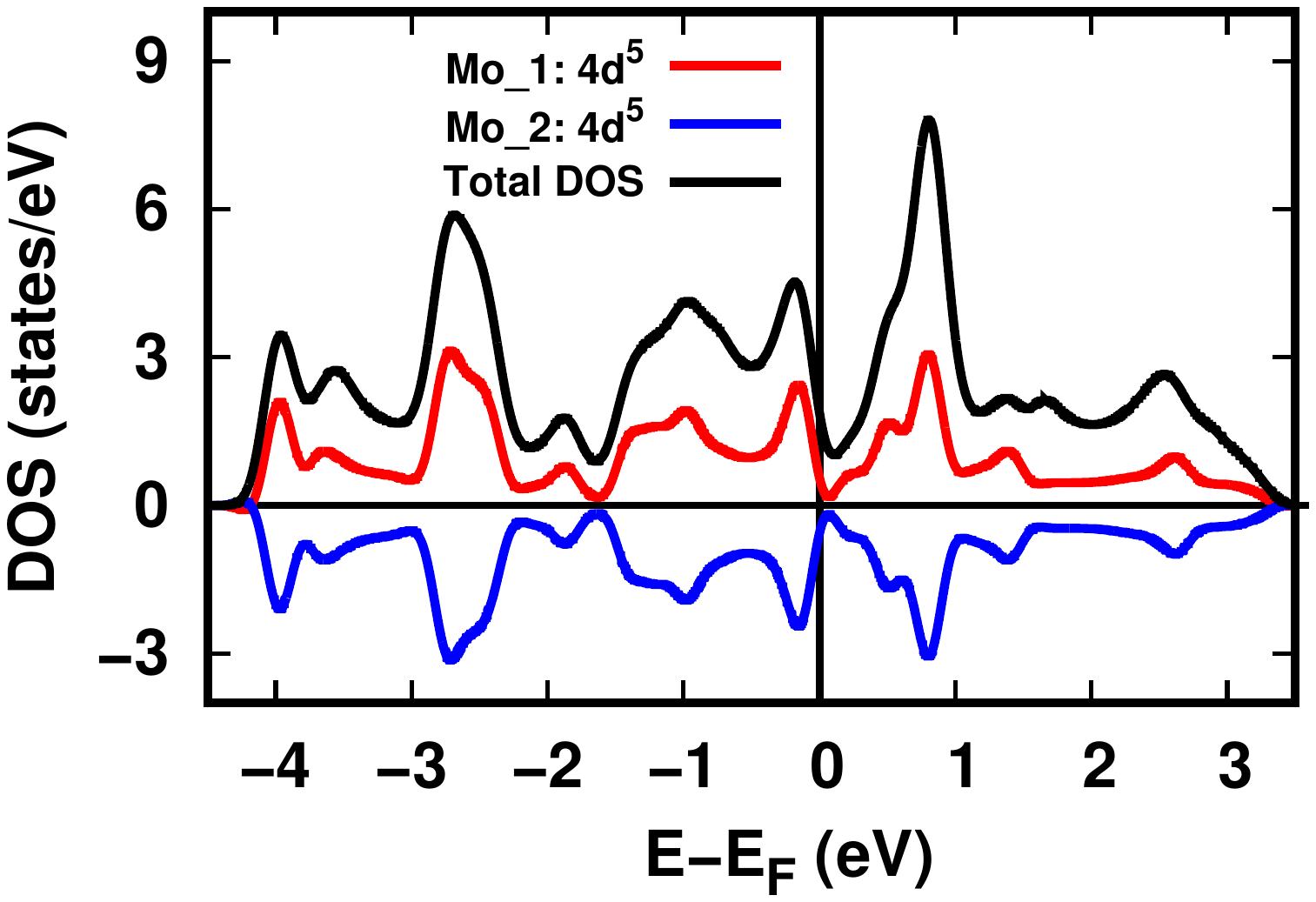}}
	\subfigure[TM]{\label{subfig:4(d)}
		\includegraphics[scale=0.29]{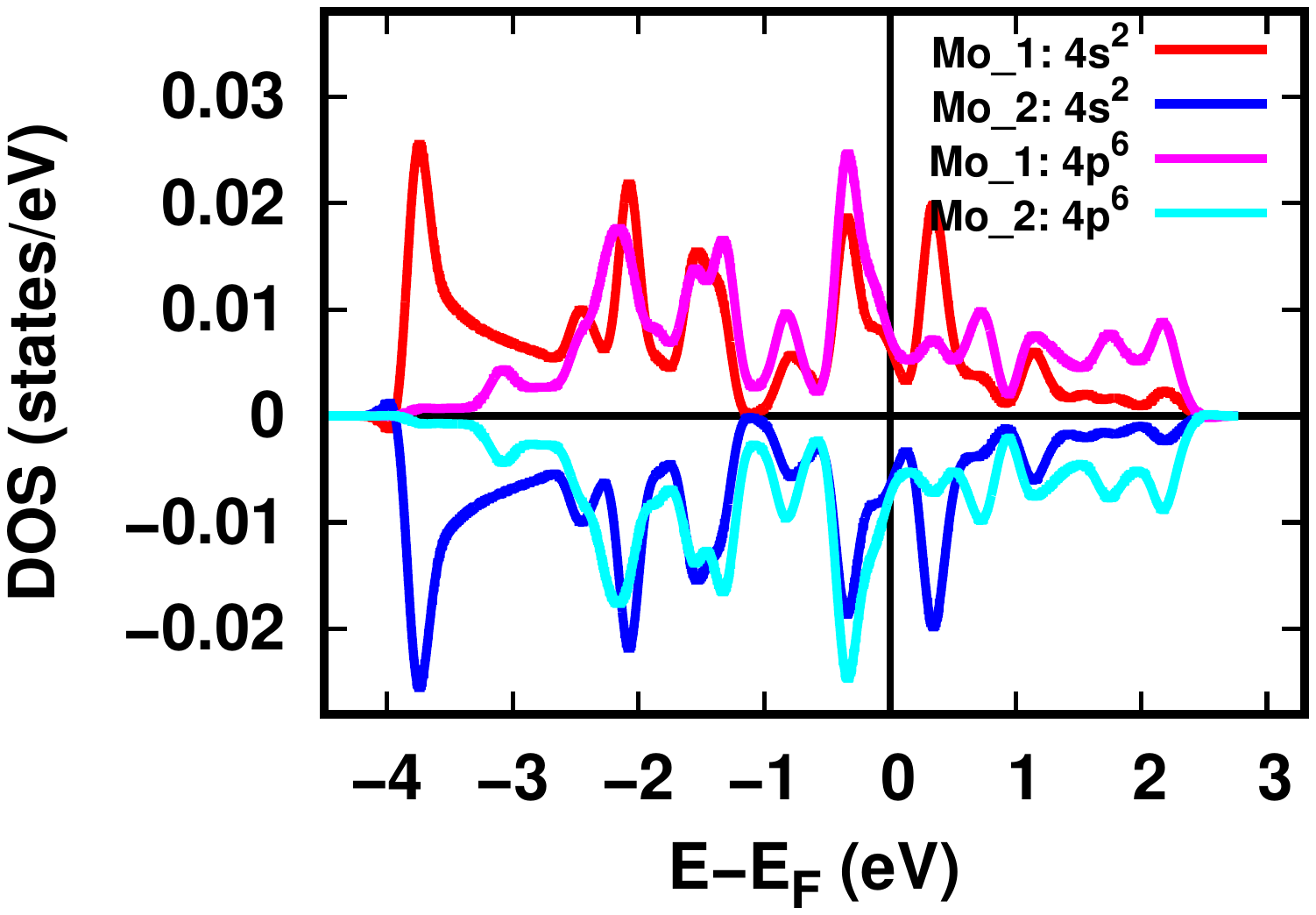}}	
	\subfigure[TM]{\label{subfig:4(e)}
		\includegraphics[scale=0.29]{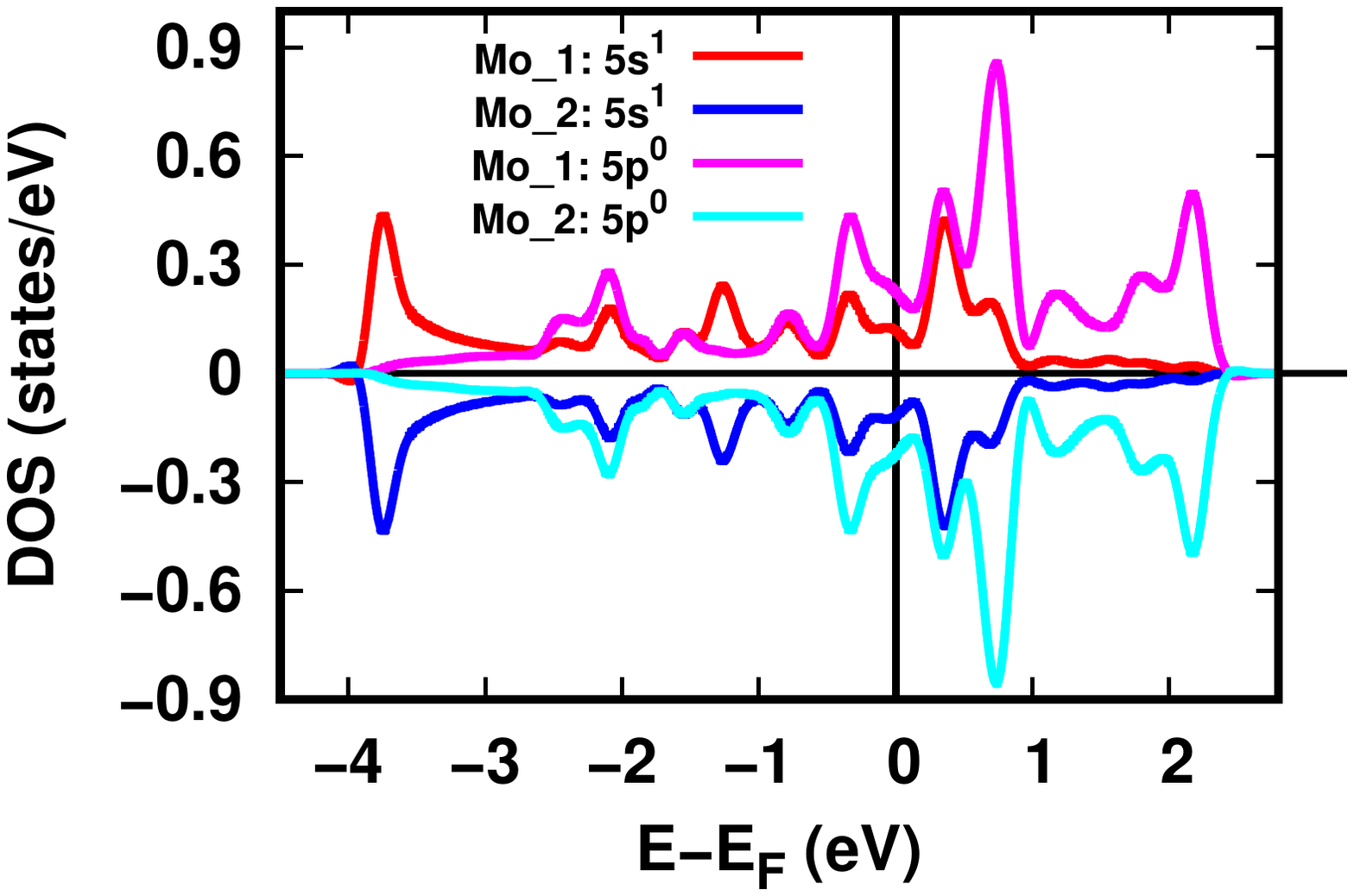}}	
	\subfigure[TM]{\label{subfig:4(f)}
		\includegraphics[scale=0.29]{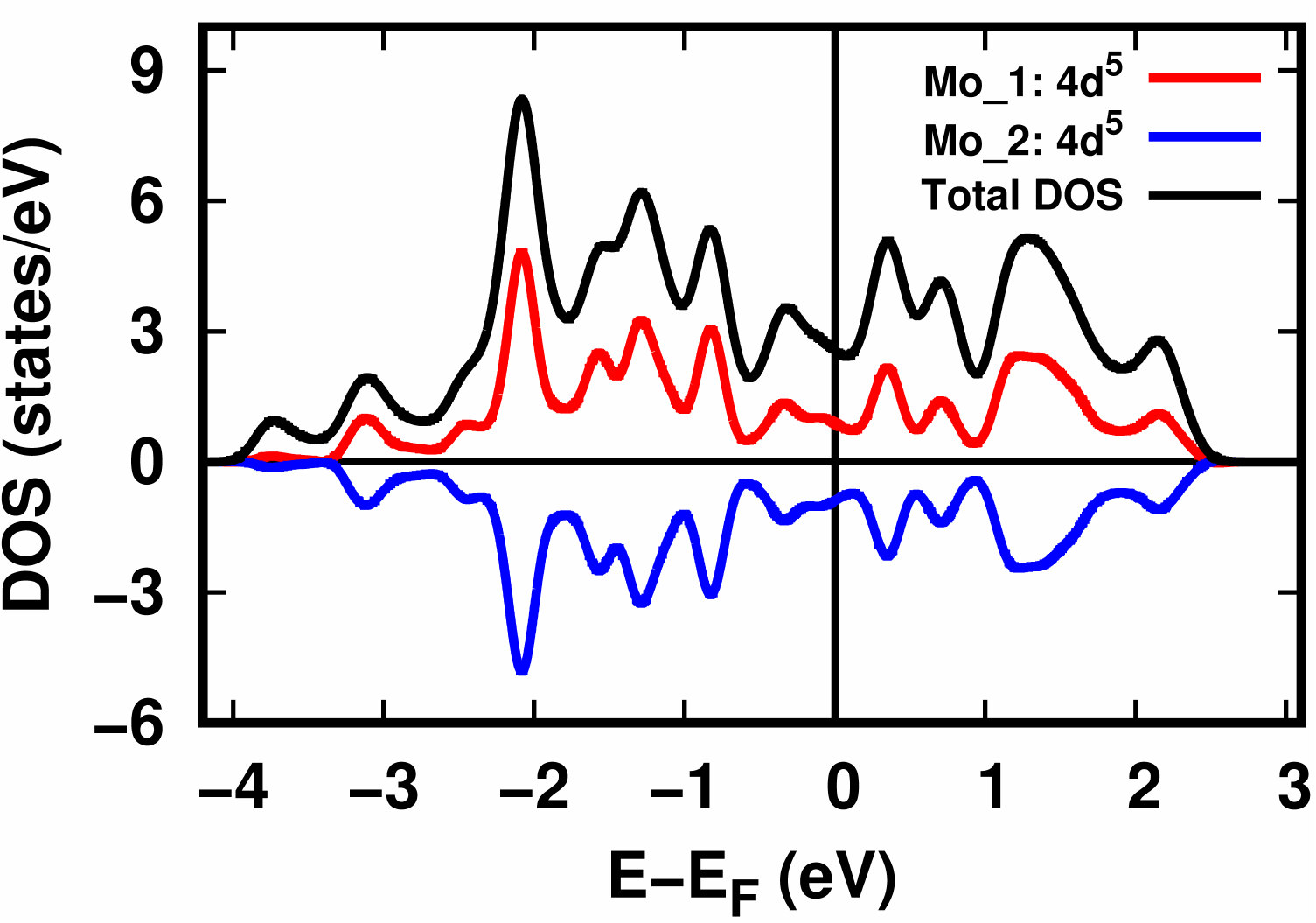}}	
	\caption{\label{fig:4}
		Electronic DOS for HM [(a) to (c)] and TM [(d) to (f)], including total DOS (black curves) as well as partial contributions of the valence orbitals $4s^2$, $4p^6$, $5s^1$, $4d^5$, and $5p^0$, of the two Mo atoms (labeled by \_1 and \_2) of the unit cells. The curves belonging to Mo\_2 have been flipped vertically to avoid overlap with those of Mo\_1. Zero of the diagrams is the Fermi level.}
\end{figure}
\end{minipage}
\end{widetext}
Electronic density of states (DOS) diagrams of Fig.~\ref{fig:4} also confirm the same result, showing total DOS as well as partial contributions of the valence orbitals $4s^2$, $4p^6$, $5s^1$, $4d^5$, and $5p^0$, for each atom of the unit cells. The metallic property of either phase is evident due to the Fermi level (zero of the diagrams) being crossed by DOS curves, consistent with the band structure analysis. According to sub-figures, the orbitals can be classified in three sets in terms of their DOS values, in ascending order, as follows: (i) $4s^24p^6$ in Figs.~\ref{subfig:4(a)} and~\ref{subfig:4(d)}; (ii) $5s^15p^0$ in Figs.~\ref{subfig:4(b)} and~\ref{subfig:4(e)}; and (iii) $4d^5$ in Figs.~\ref{subfig:4(c)} and~\ref{subfig:4(f)}.\\

From Figs.~\ref{subfig:4(a)} and~\ref{subfig:4(d)}, the orbitals $4s^2$ and $4p^6$ have the least contribution to electric conductivity based on the fact that these orbitals have no empty states; therefore according to Pauli's exclusion principle, electrons cannot enter$/$move within such full states.\\

According to Figs.~\ref{subfig:4(b)} and~\ref{subfig:4(e)}, the orbitals $5s^1$ and $5p^0$ have larger contributions compared to set (i), based on the same reasoning: half of the former ($5s^1$) and entire of the latter orbital ($5p^0$) are empty, giving charge carriers within$/$adjacent to these orbitals the required kinetic degree of freedom to conduct electricity as well. Since electronic configuration of Mo is [Kr] $5s^14d^5$, the next orbital $5p$ has accordingly no atomic electron (so, $5p^0$); therefore, only conduction electrons can enter this orbital according to Drude's model~\cite{drd,ashc}.\\

The largest contributions to total DOS, according to Figs.~\ref{subfig:4(c)} and~\ref{subfig:4(f)}, arise from $4d^5$ orbitals, which are the outermost orbitals (then contain weakest-bound atomic electrons) the halves of which are empty. They accordingly lead to largest share in electric conductivity of either phase. This is why the shape of the total DOS (black curves) is mainly determined by $4d^5$.
\subsection{Lattice dynamics}
To investigate the lattice stability of either molybdenene monolayer, we also have applied phonon-dispersion calculations along the aforementioned k-space paths described in Sec.~\ref{sec:2}, the results of which have been illustrated in Fig.~\ref{fig:5} in the form of phonon band and DOS diagrams.
\begin{widetext}
\begin{minipage}{\linewidth} 
\begin{figure}[H]
	\centering
	\subfigure[HM]{\label{subfig:5(a)}
		\includegraphics[scale=0.29]{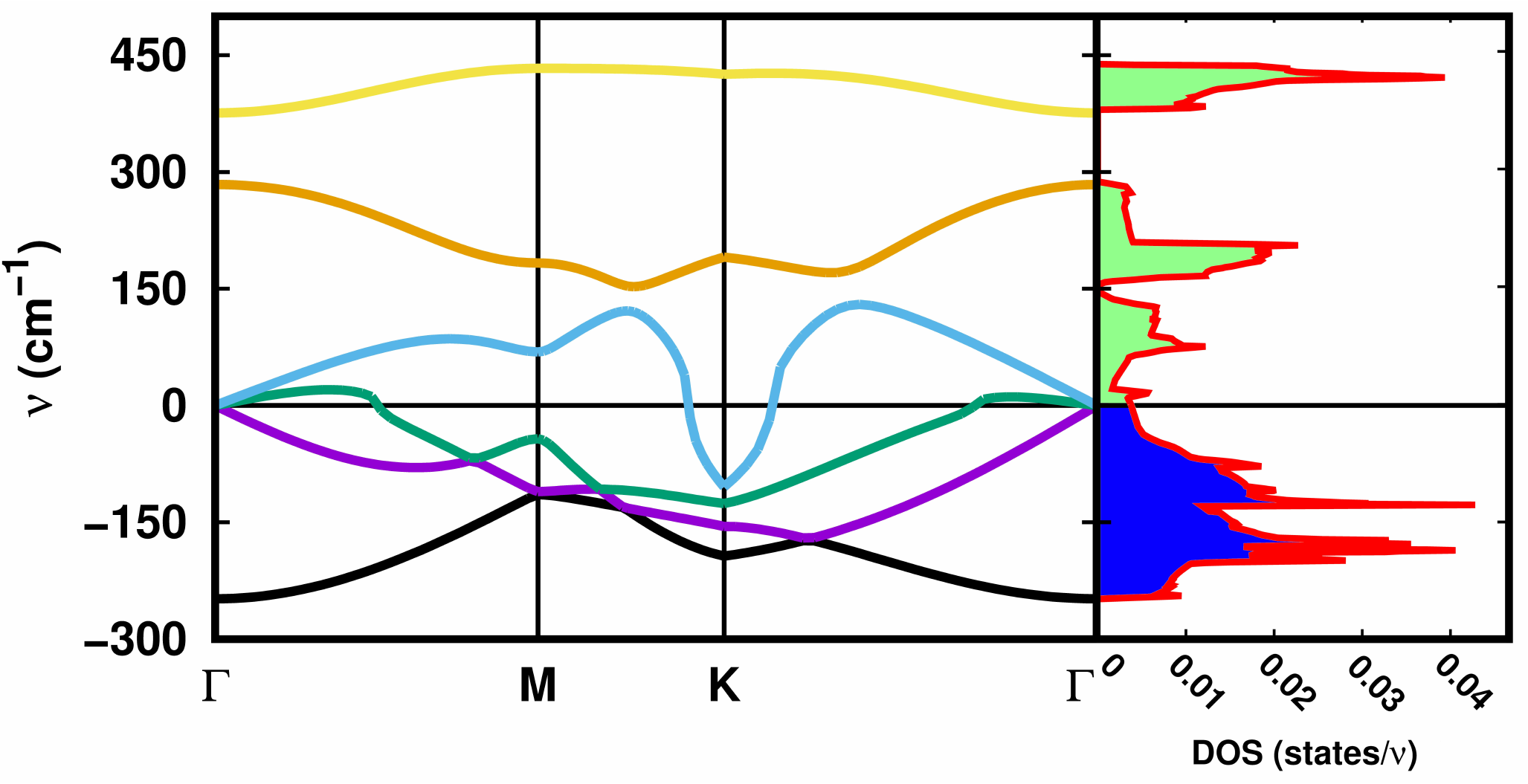}}
	\subfigure[TM]{\label{subfig:5(b)}
		\includegraphics[scale=0.29]{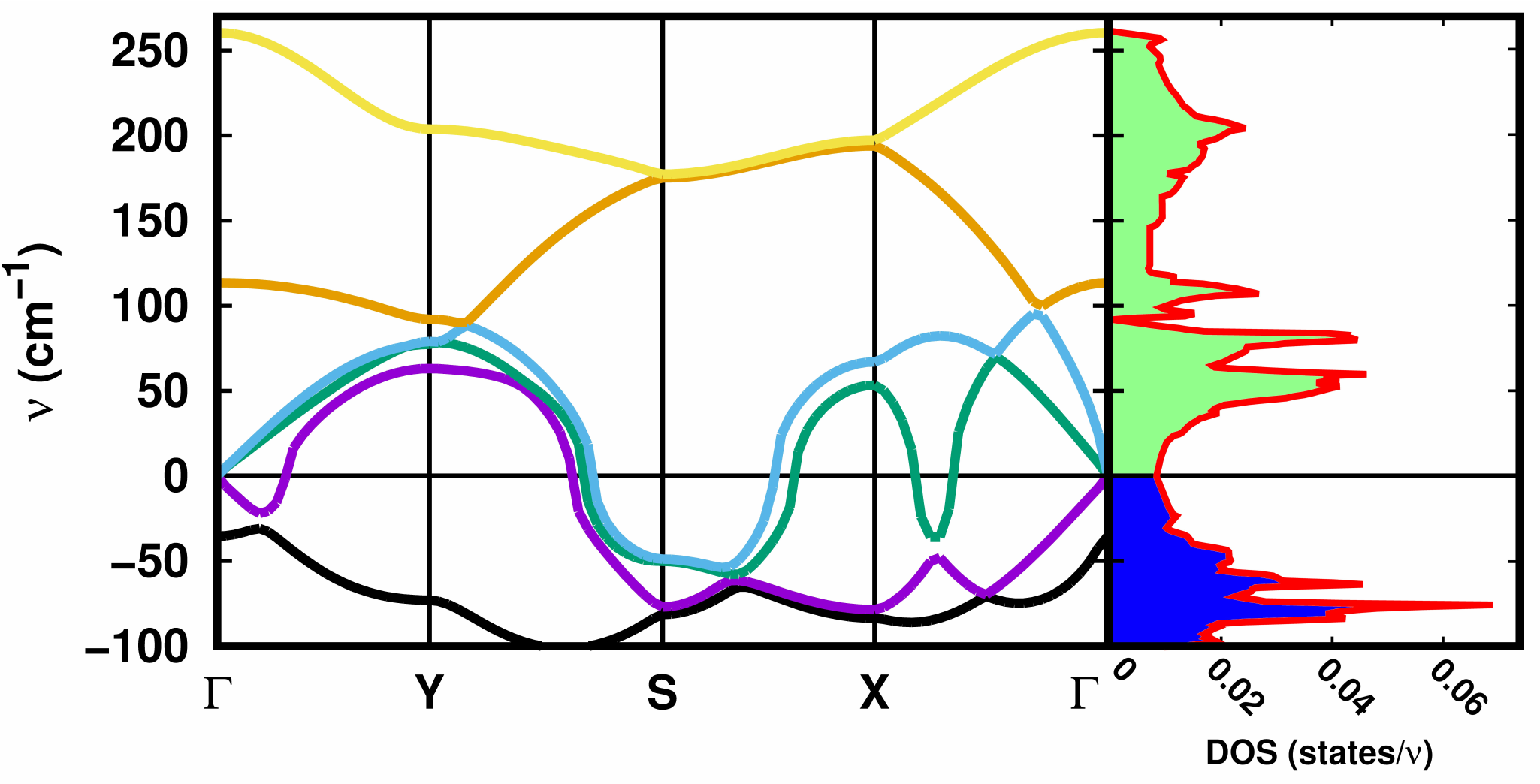}}	
	\caption{\label{fig:5}
		Phonon band structure and DOS for (a) HM and (b) TM. Negative frequencies or soft modes (corresponding to the blue areas of DOS) are clear manifestations of lattice instability, observed in both phases.}
\end{figure}
\end{minipage}
\end{widetext}
As is seen, there are negative frequencies known as soft modes over the entire k-space paths, showing that either phase is unstable in terms of lattice dynamics.
\section{\label{sec:4}Conclusions}
The ultrasoft pseudopotential approach within the framework of density functional theory (DFT) has been applied to inquire into the crystal structure, electronic properties, and lattice dynamics of molybdenene monolayer in both its hexagonal (HM) and triclinic (TM) phases. For either monolayer, the equilibrium, zero-pressure values of lattice constant and bulk modulus have been found via the use of Murnaghan isothermal equation of state and by examining the related isothermal $P-V$ and $E-V$ diagrams. In agreement with experimental findings, either phase has showed metallic properties according to its electronic band structure and density of states (DOS) diagrams. We classified the partial DOS diagrams into three sets in terms of their numerical values, which are, in ascending order, $4s^24p^6$, $5s^15p^0$, and $4d^5$. They further revealed that the $4d^5$ valence orbitals of the Mo atoms of the unit cells have had the largest contribution to such a metallic feature based on the facts that such orbitals are not only half-empty, but also the outermost shells. Phonon-dispersion calculations also led to negative frequencies (soft modes) for both monolayers, showing that they are unstable in terms of lattice dynamics.
\section*{Data availability}
The authors confirm that the data supporting the findings of this investigation are available within the present article. The associated raw data are also available upon request from the corresponding author.
\section*{Competing interests}
The authors declare no competing interests.
\section*{Additional information}
{\bf{Correspondence}}  should be addressed to M. Jafari.

\end{document}